\documentclass[letterpaper, 10 pt, conference]{ieeeconf}  
\IEEEoverridecommandlockouts                              
\usepackage{graphicx}      
\usepackage{balance}
\usepackage{pdfpages}  
\usepackage{float}
\usepackage{algorithmic}
\usepackage{algorithm}
\usepackage{amssymb, psfrag, units, booktabs}
\usepackage{amsmath}
\usepackage{pgfplots}
\usepackage{eurosym, color, subfloat, xfrac}
\usepackage{placeins}
\usepackage{tikz}
\usetikzlibrary{shapes,arrows}
\usepackage{circuitikz}
\usepackage{bytefield}
\usepackage{relsize}
\usetikzlibrary{decorations.pathreplacing}
\usepackage{adjustbox}
\usepackage{float}
\usepackage{siunitx}
\usepackage{multirow}
\usepackage{fancyvrb}

\newcommand{\lrp}[1]{\ensuremath{\left( #1 \right)}}
\makeatletter
\let\AND\original@AND
\makeatother
\newcommand{\matlab}{\textsc{matlab}\textsuperscript{\textregistered}}
\newcommand{\mplab}{\textsc{mplab}\textsuperscript{\textregistered}}

\newcommand{\pic}{PIC~$18$F$4550$}
\newcommand{\SIM}{\textsc{matlab/simulink}}

\newcommand\cp{C\nolinebreak[4]\hspace{-.05em}\raisebox{.4ex}{\relsize{-3}{\textbf{}}}}
 
\newcommand{\kp}{k_p}
\newcommand{\ki}{k_i}
\newcommand{\ts}{T_s}
\title{\LARGE \bf Embedded Model Predictive Controller on Low-Cost Low-End Microcontroller for Electrical Drives}
\author{Saket Adhau$^{1}$, Siddesh Dani$^{2}$, Deepak Ingole$^{3}$, and Dayaram Sonawane$^{1}$
	\thanks{$^{1}$ Department of Instrumentation and Control Engineering, College of Engineering, Pune, India, {\tt\small adhauss17.instru@coep.ac.in, dns.instru@coep.ac.in.}}
	\thanks{$^{2}$ Renu Electronics Private Limited, Pune, India, {\tt\small danisk15.instru@coep.ac.in}.}
	\thanks{$^{3}$ University of Lyon, IFSTTAR, ENTPE, Lyon, France, {\tt\small deepak.ingole@ifsttar.fr}.}
}
\begin{document}
	\maketitle
	\thispagestyle{empty}
	\pagestyle{empty}
\begin{abstract}
It is very well-known that the implementation of Model Predictive Controller (MPC) on embedded platforms is challenging due to the computational complexities associated while solving an optimization problem. Although, there are many efficient embedded implementations existing by now, but for faster, more dynamic and non-linear control applications, there is no cost-effective and memory efficient embedded solutions. In this paper, we show the implementation of embedded explicit MPC for a motor speed control application on a low-cost \SI{8}{bit} PIC 18 series microcontroller which costs only $\$ 5$. The offset-free explicit MPC is designed for reference tracking, constraints handling, and disturbance rejection. The developed control law is exported to low-level~\cp~code and utilized in HIL co-simulations. We present the results of memory demand and control performance under various operating scenarios. The presented results show that the developed embedded MPC utilize about 40$\%$ of RAM and 92$\%$ of ROM for prediction horizon up to 3 samples. The performance of developed MPC is compared with the conventional PI controller. Overall results show that the presented approach is cost-effective, portable, and gives better performance than the PI controller. 
\end{abstract}

\begin{keywords}
		motor control, model predictive control, disturbance modeling, microcontroller, optimization.
\end{keywords}
\section{INTRODUCTION}
Looking at the performance of Model Predictive Controller (MPC), along with the better constraint handling, offset-free reference tracking and optimal control actions prove the best choice in many process industries and extended applications. There are many different ways to implement MPC on embedded platforms for standalone applications. MPC has also been implemented on System on Chip (SoC) by~\cite{bleris2005real} which presents a basic framework and hardware architecture required for real-time MPC. 
Authors in~\cite{vouzis2009system} have developed MPC for micro-chemical systems using Motorola's MPC$555$ which has high speed~\SI{32}{bit} CPU and special~\SI{64}{bit} floating-point unit designed to accelerate advanced algorithms for complex applications. Many different embedded architectures for linear MPC have been presented in~\cite{frison2014efficient}. Field Programmable Gate Array (FPGA) is another hardware tool which has been extensively used for implementing MPC~\cite{yang2012model}. FPGA requires more coding efforts and are complex in prototyping. Also, they consume a lot more power than any other~\cite{kheriji2011microcontroller}. So, research is being carried out to implement MPC on microcontrollers which are less expensive and consume less power than FPGAs. On the similar line, Generalized Predictive Control (GPC) on an embedded platform (STM$32$) has been proposed by the authors in~\cite{kheriji2011microcontroller}. 
The authors in~\cite{abbes2011microcontroller} have implemented constrained MPC for the first order and second order plant on the STM$32$ kit and have compared the performance with anti-windup Proportional-Integral-Derivative (PID). 

Based on the optimization problem-solving approach MPC can be categorized in two ways, namely on-line MPC and off-line or Explicit MPC (EMPC). On-line MPC is computationally burdening on the processor and require a longer time for solving quadratic problems. Therefore, it requires processor working at higher clock speeds. On the other hand, explicit MPC comprised of some data structure in the form of Look-Up Tables (LUTs) and point location algorithm which efficiently searches through for current optimal MPC solution. Hence, it is computationally less complex and therefore less computationally burdening on the processor~\cite{johansen2015toward}. On the downside, the memory requirement for EMPC is high as compared to on-line MPC and it increases with the horizons. Also, there are many EMPC solutions available on embedded platforms. Authors in~\cite{ingole2015fpga,ingole2015fpgabst} has proposed an FPGA-based EMPC scheme using Binary Search Tree (BST) to find the optimal rate of anesthesia drug infusion in the patient. 
The authors in~\cite{takacs2016embedded} have presented a real-time application of EMPC for active vibration compression using a~\SI{32}{bit} ARM Cortex M$4$F (STM$32$F$407$VGT$6$). 

Motivated by these research ideas and our previous work on~\matlab~implementation of on-line MPC~\cite{dani2017performance}. This paper attempts to implement EMPC solution with disturbance modeling on low-cost microcontroller for speed control of DC motor. The sole reason for selecting this low-cost microcontroller is to prove that the explicit MPC can be implemented on low-end microcontroller for dedicated standalone applications. An offset-free explicit MPC is designed for the Multi-Input Single-Output (MISO) motor model using disturbance modeling approach. Furthermore, a low-level~\cp~code of the explicit MPC is exported and deployed on the microcontroller which is then used to control the desired speed reference using Hardware-in-the-Loop (HIL) co-simulation. The obtained results of the designed explicit MPC are presented for reference tracking, disturbance rejection, and constraint handling along with the results of PI controller.
\section{DC Motor Modeling}
\label{sec:pltmodel}
\begin{figure}[htbp]
		\centering
		\newlength\figureheight \newlength\figurewidth 
		\setlength\figureheight{5cm} \setlength\figurewidth{0.9\textwidth} 
			\begin{circuitikz}[american voltages]
		\draw (0,3) to[V, l_ = $V_a$] (0,0);
		\draw (0,3) to[R, i>^=$i_a$, l=$R_a$] (3,3);
		\draw (3,3) to[L, l=$L_a$] (4,3);
		\draw (4,3) -- (5,3);
		\draw (5,0) to [V] (5,3) ;
		\draw (0,0) -- (5,0);
		\draw[fill=black] (4.85,0.85) rectangle (5.15,2.15);
		\draw[fill=white] (5,1.5) ellipse (.45 and .45);
		\draw[fill=black] (5.45,1.45) rectangle (6.5,1.55);
	  	\draw[line width=0.7pt,<-] (5.8,1) arc (-30:30:1);
		\draw[fill=white] (6.5,1.55)ellipse (.15 and 0.4);
		\draw (7.5,1.55) ellipse (.15 and 0.4);
		\draw (6.5,1.95) -- (7.5,1.95);
		\draw (6.5,1.15) -- (7.5,1.15);
		\draw (5.85,2.2) node {$\omega_m, \theta_m$};
		\draw (5.85,0.7) node {$T_m$};
		\draw (7,0.9) node {$J, f_m$};
		\draw (5,1.5) node {M};
		\draw (4.1,1.5) node {$V_{m}$};
	\end{circuitikz}
	
		\caption{Electrical equivalent circuit of DC motor.} 
		\label{fig:electrical:ckt}
	\end{figure}
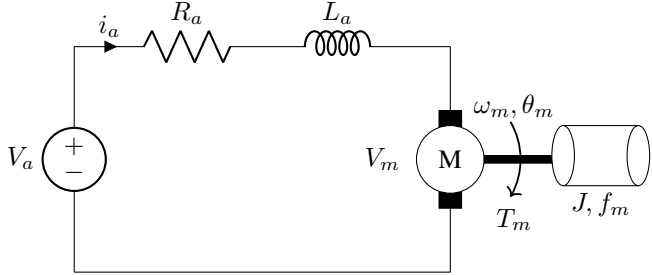

In this section, a mathematical model of permanent magnet DC motor is presented. The model is derived using torque and electrical characteristics described in~\cite[Chapter 2]{krishnan2001electric}. The electrical circuit of the motor is shown in Fig.~\ref{fig:electrical:ckt}. A voltage source ($V_a$) across the armature coil, induced voltage or back electromotive force (emf) ($V_m$) which opposes the voltage source  and an inductance ($L_a$) in series with a resistance ($R_a$)  forms the closed circuit. The rotation of electrical coil through fixed flux lines of permanent magnets generates the back emf.  
Differential equations for the electrical circuit shown in Fig.~\ref{fig:electrical:ckt} are derived using Kirchoff's voltage law. The sum of all voltages around a loop can be given using Ohm's and Kirchoff's law
	\begin{equation}
	\label{eq:model:kvl}
	L_{a}\frac{di_a}{dt}+R_{a}i_{a}+V_{m}=V_{a},
	\end{equation} 
where $\frac{di_a}{dt}$ is the change in current through the coil with respect to time and $i_a$ is the armature current. The back emf is obtained by 
	\begin{equation}
	\label{eq:model:emf}
	V_m = k_m \omega_m,
	\end{equation}
where $\omega_m$ is the rotational speed of the rotor and $k_m$ is the back emf constant described by the reluctance of the iron core of armature, flux density of permanent magnets and the number of turns in the armature winding.
	
The applied voltages at stator and rotor causes the motor to exerts a torque. This torque acting on the  mechanical shaft  is characterized by the viscous friction coefficient $f_m$ and rotor inertia $J$. If external disturbances are acting on the shaft of the motor categorized as load torque ; then,  $T_l$ is load torque and $T_m = k_mi_a$ is emf torque, the following equation can be written for the total torque of motor as
	\begin{equation}
	\label{eq:model:torque}
	J\frac{d\omega_m}{dt}+f_{m}\omega_m=k_{m}i_{a} - T_l.
	\end{equation}
Applying the Laplace transform to motor model~\eqref{eq:model:kvl} and~\eqref{eq:model:torque}, in terms of the Laplace variable $s$ we get
	\begin{align}
	\label{eq:model:lap:kvl}
	(L_a s + R_a)i_a(s)+ k_ms\omega_m(s)&= V_a(s), \\
	\label{eq:model:lap:torque}
	s(Js+f_m)\omega_m(s) &= k_mi_a(s).
	\end{align}
Considering the armature voltage ($V_a(s)$) as the input variable, the rotational speed ($\omega_m(s)$) as the output variable and by eliminating $i_a(s)$ from~\eqref{eq:model:lap:kvl} and~\eqref{eq:model:lap:torque}, we arrive at the following open-loop transfer function,	
	\begin{equation}
	\label{eq:model:tf:1}
	G_{\omega_m V_a}(s)= \frac{k_m}{(R_a+sL_a) (Js+ f_m) + k_m^2}.
	\end{equation} 		
For linearizing the system and as $L_{a} << R_{a}$,  $L_{a}$ term in~\eqref{eq:model:tf:1} is neglected. We get the new transfer function as,
	\begin{equation}
	\label{eq:model:tf:2}
	G_{\omega_m V_a}(s)=\frac{k_m}{R_a(Js + f_m) + k_m^2}.
	\end{equation}
The viscous friction coefficient ($f$) of motor can be stated as
	\begin{equation} 
	\label{eq:model:vis}
	f=f_m+k_m^2/R_{a}.
	\end{equation}	
Further simplification gives,
	\begin{equation}
	\label{eq:model:tf:3}
	G_{\omega_m V_a}(s)=\frac{K}{(\tau s+1)},
	\end{equation}	
where $K$ is steady state gain $\frac{k_{m}}{R_{a}f}$ and $\tau$ is the  time constant of the system described as $\frac{J}{f}$. Similarly  the transfer function of the angular position ($\theta_m$) can be obtained by integrating angular speed i.e. multiplying $\frac{1}{s}$ to~\eqref{eq:model:tf:2} as 
	\begin{equation}
	\label{eq:model:tf:4}
	G_{\theta_m V_a}(s)=\frac{K}{s(\tau s+1)}.
	\end{equation}
Here, angular position and angular velocity as considered as the states and armature voltage as the input, the equation in state space form becomes, 
	\begin{subequations}
		\label{eq:clti}
		\begin{align}
		\label{eq:clti:state}
		\dot x(t)&=A x(t) + B u(t),\\
		\label{eq:clti:output}
		y(t)&=C x(t) + D u(t),
		\end{align}
	\end{subequations}
where system matrices $A$, $B$, $C$ and $D$ are given as
	\begin{align*}
	A=\begin{bmatrix} 0 &1\\0 &-\frac{1}{\tau} \end{bmatrix},
	B=\left[ \begin{array}{c} 0 \\\frac{K}{\tau}  \end{array} \right],
	C=\left[ \begin{array}{c} 0 \\ 1 \end{array} \right]^{T},
	D=0.
	\end{align*}
Table~\ref{tab:1} shows the system parameters for simulation of the model.	
	\begin{table}[htbp]
		\centering
		\setlength{\tabcolsep}{3pt}
		\renewcommand*{\arraystretch}{1.3}
		\caption{System Parameters.}
		\begin{tabular}{llll}
			\toprule
			Notation &  Description & Value &  Unit\\\midrule
			$k_{m}$& Torque constant& $8.32\times 10^{-4}$ &\unit{Nm/A}\\ 
			$J$& Moment of inertia& $2.45\times10^{-7}$&\unit{kg.m$^{2}$}\\ 
			$f_{m}$& Viscous friction constant&$3.10\times 10^{-5}$&\unit{Nm/rad/s}\\ 
			$R_{a}$& Armature resistance&$4.1$&\unit{$\Omega$}\\ 
			$L_{a}$& Armature inductance&$2.27$&\unit{mH}\\ 
			\bottomrule
		\end{tabular}
		\label{tab:1}
	\end{table}
In the next, an overview of designed control algorithms is given.
\section{Proportional Integral (PI) Controller}
\label{sec:PI}
	
PI controller algorithm consists of two blocks namely, the Proportional and Integral mode. 
The simplest form of the PI controller in continuous-time is given by
\begin{eqnarray}
\label{eq:pi:input:ct}
u_t=\kp e_t+\ki\int_{0}^{t} e_{\tau} d\tau,
\end{eqnarray}
where $u$ is the control input, $e$ is the error between reference and output, and $\kp$ and $\ki$ are the proportional and integral gain, respectively. The integral action removes the offset in the error and the  proportional action is related to the increase of the control variable when the control error is large. The discrete time equation is:
\begin{eqnarray}
\label{eq:pi:input:dt}
u_k=\kp e_k+\ki \ts \sum_{j=0}^{k}{e_j},
\end{eqnarray}
where $\ts$ is the sampling time of the system.
\section{Offset-Free Explicit MPC}
\label{sec:MPC}
\subsection{Plant Model}
Consider a discrete-time version of the Linear-Time Invariant (LTI) system in~\eqref{eq:clti},
	\begin{subequations}
		\label{eq:dlti}
		\begin{align}
		\label{eq:dlti:state}
		x(t+T_s) &= Ax(t)+Bu(t),\\
		\label{eq:dlti:output}
		y(t)&= Cx(t) + Du(t),
		\end{align}
	\end{subequations}
where $x(t)\in \mathbb{R}^{n}$ is the system state vector, $u(t)\in \mathbb{R}^{l}$ is the system input vector and $y(t)\in \mathbb{R}^{m}$ is the system output vector, moreover, $A \in \mathbb{R}^{n\times n}$, $B\in\mathbb{R}^{n\times l}$, $C\in\mathbb{R}^{m\times n}$ and $D \in \mathbb{R}^{m \times l}$ are system matrices.
\subsection{Disturbance Modeling}
The objective is to design an explicit model predictive controller based on linear system model~\eqref{eq:clti} in order to have measured output $y(t)$ track the desired reference $r(t)$ with zero steady-state error in presence of the plant-model mismatch and/or un-known disturbances. To achieve this objective, the plant model~\eqref{eq:dlti} is augmented with a disturbance vector $d(t) \in \mathbb{R}^p$~\cite{pannocchia2003disturbance},~\cite[Chapter 12]{borrelli2017predictive} as shown below in state space form
	\begin{subequations}
		\label{eq:alti}
		\begin{align}
		\underbrace{{\left[\begin{array}{c}{x}(t+T_s) \\ {d}(t+T_s)\end{array} \right]}}_{x_e(t+T_s)} & = \underbrace{\begin{bmatrix} A & B_d \\ 0 & I \end{bmatrix}}_{A_e}\underbrace{{\left[\begin{array}{c} {x}(t) \\ {d}(t) \end{array} \right]}}_{x_e(t)} + \underbrace{{\left[\begin{array}{c}{B} \\ 0\end{array} \right]}}_{B_e} u(t), \\
		y_e(t)  & = \;\underbrace{\begin{bmatrix} C & C_d \end{bmatrix}}_{C_e}\;\underbrace{{\left[\begin{array}{c} {x}(t)  \\ {d}(t) \end{array} \right]}}_{x_e(t)} + D_eu(t),
		\end{align}
	\end{subequations}
where $B_d \in \mathbb{R}^{n\times p}$, $C_d\in \mathbb{R}^{m\times p}$ are the disturbance model matrices and dimensions of matrices $0$ and $I$ are ${m\times l}$ and ${m\times m}$ respectively. The subscript `$e$' is the extended version of the combined disturbance and state.
\subsection{State and Disturbance Estimation}
Extended state $x_e$ is estimated from the plant measurement by designing a Luenberger observer for augmented system~\eqref{eq:alti} as follows,
	\begin{subequations}
		\label{eq:esti}
		\begin{align}
		\label{eq:esti:state}
		\hat{x}_e(t+T_s) &= A_e \hat{x}_e(t) + B_e u(t) + L_e (y(t) - \hat{y}_e(t)),\\
		\label{eq:esti:output}
		\hat{y}_e(t) & = C_e\hat{x}_e(t) +D_eu(t),
		\end{align}
	\end{subequations}	
where $L_e = \left[\begin{smallmatrix} L_x & L_d \end{smallmatrix}\right]^T$ is the filter gain matrices for the state (of dimension $n \times m$) and the disturbance (of dimension $p \times m$), respectively and can be obtained by pole placement.
\subsection{MPC Formulation}
\label{subsec:FormulMPC}
In MPC, the control action is obtained by solving a Constrained Finite Time Optimal Control (CFTOC) problem for the current state ($\hat{x}_e(t)$) of the plant at each sampling time $(t)$. The sequence of optimal control inputs ($U^{\star} = \{u_0^{\star}, \ldots, u_{N-1}^{\star}\}$) is computed for a predicted evolution of the system model over a finite horizon (or prediction horizon ($N$)). However, only the first element of the control sequence ($u_0^{\star}$) is applied and the current state ($\hat{x}_e(t)$) of the system is then measured again at the next sampling time ($t+1$). This so-called Receding Horizon Controller (RHC) which introduces feedback to the system, thereby allowing for the compensation of potential modeling errors or disturbances acting on the system~\cite[Chapter 12]{borrelli2017predictive}. 
	
Using LTI system model in~\eqref{eq:dlti} and disturbance observer in~\eqref{eq:esti} the MPC problem is designed as follows 
\begin{subequations}
	\label{eq:cftoc}
	\begin{align}
	\label{eq:cftoc:obj}
	&\hspace{-1.47cm}\min _{U} \sum _{k = 0} ^{N-1} \lrp{y_k - r_k}^TQ \lrp{y_k - r_k} + \Delta u_k^T R \Delta u_k \\
	\label{eq:cftoc:state}
	\text{s.t.} \;\; x_{{k+1}} &= Ax_{k} + Bu_{k} + B_d d_{k},\\ \label{eq:cftoc:dist}
	d_{k+1} & = d_{k}, 
	\\ \label{eq:cftoc:output}
	y_{k} & = Cx_{k} + Du_{k} + C_dd_{k},\\ \label{eq:cftoc:deltau}
	\Delta u_k &= u_k - u_{k-1},\\
	\label{eq:cftoc:const:u}
	u_{\min} &\leq u_{k} \leq u_{\max}, \\
	\label{eq:cftoc:ini:u}
	u_{-1} &= 0, \\ 	\label{eq:cftoc:ini:state}
	x_0 &= \hat{x}(t),\\\label{eq:cftoc:ini:dist}
	d_0 &= \hat{d}(t),\\
	\forall k &\in\{k = 0, \ldots, N-1\},
	\end{align}
\end{subequations}
where $Q \in \mathbb{R}^{n \times n}$ and $R \in \mathbb{R}^{l \times l}$ are the weighting matrices, with condition $Q \succeq 0$ to be positive semi-definite, and $R \succ 0$ to be positive definite. We denote $N$ as the prediction horizon, $x_{k+1}$ as the vector of predicted states at sample time $k$, $U = \{u_0, \ldots, u_{N-1}\}$ as the sequence of control actions, $r_{k}$ is the output reference trajectory to be tracked, and $x_0$, $d_0$, and $u_{-1}$ are the given initial conditions.
	
By solving~\eqref{eq:cftoc} with a given initial conditions, the optimization yields open loop optimal input sequence $U^{\star} = \{u_0^{\star}, \ldots, u_{N-1}^{\star}\}$ from which only the first control action, i.e., $u_0^{\star}$ is applied to the plant and again a CFTOC problem~\eqref{eq:cftoc} is solved at next time instance $k+1$. 
\subsection{Explicit MPC}
\label{subsec:empc}
	
In MPC, an optimization problem~\eqref{eq:cftoc} needs to be solved at each time instance. Such an optimization problem can be formulated as a multi-parametric Quadratic Programming (mp-QP) problem 
	\begin{subequations}
		\label{eq:mpqp}
		\begin{align}
		\underset{U}{\text{min}} & \; \bigg\{U^THU + \tilde{x}_{0}^TFU\bigg\} + \tilde{x}_{0}^TY\tilde{x}_{0},\label{eq:mpqp:cost}\\
		\text{s.t.} \; 	& \; G U \le w + W \tilde{x}_{0}, \label{eq:mpqp:const}
		\end{align}
	\end{subequations}
where $\tilde{x}_{0} = [\hat{x}_e^T(t)~u^T(t-1)~r^T(t)]^T$ is the vector of initial conditions and by denoting $q$ as a number of inequalities, matrices $H \in \mathbb{R}^{l.N \times l.N}, F \in \mathbb{R}^{(n+p) \times (l.N)}, Y \in \mathbb{R}^{(n+p) \times (n+p)}, G \in \mathbb{R}^{q\times l.N}, w \in \mathbb{R}^{q}, W \in \mathbb{R}^{q \times n}$ can be obtained by weighting matrices $Q$ and $R$~\cite{bemporad2002explicit}.
	
The optimal solution $U^{\star}$ is a piecewise affine function of the initial condition, which can be computed off-line by solving mp-QP problem~\cite{bemporad2002explicit}. This mp-QP solution can be evaluated using~\matlab~based Multi-Parametric Toolbox~(MPT)~\cite{herceg2013mpt,kvasnica2015design}. Once the mp-QP problem~\eqref{eq:mpqp} is solved off-line and stored locally in LUTs, explicit MPC uses the obtained optimal solution in a receding horizon fashion in which $U^{\star} = \kappa(\tilde{x}_{0})$ is a continuous PWA function mapped over a polyhedral partition
	\begin{equation}
	\label{eq:pwa}
	\kappa(\tilde{x}_{0}) = \begin{cases}
	F_1\tilde{x}_{0}+g_1  & \text{if} \ \tilde{x}_{0} \in \mathcal{R}_1\\
	& \vdots\\
	F_M\tilde{x}_{0}+g_M  & \text{if} \ \tilde{x}_{0} \in \mathcal{R}_{M}
	\end{cases}    
	\end{equation}
where $\mathcal{R}_i = \{\tilde{x} \in\mathbb{R}^{(n+p)} \ | \ Z_i \tilde{x} \le z_i \} \;\; \forall \, i = 1, \dots, M$ are the polyhedral regions and $F_i \in \mathbb{R}^{l\times (n+p)}$, $g_i \in \mathbb{R}^{l}$ are optimal gains. Also, $M$ states the total regions, $h_i$ is a number of half-spaces in a polyhedral set and $Z_i\in\mathbb{R}^{h_i\times(n + p)} $, $z_i\in \mathbb{R}^{h_i}$ are the matrices which forms a polyhedral set. The advantage of explicit form of controller as shown in~\eqref{eq:pwa} is, the computation is reduced to only search algorithm consisting of addition and multiplications to find the optimal control inputs . 
\subsection{Evaluation of PWA Function}
\label{subsec:plp}
In the on-line phase of explicit MPC, the task is to identify an index of polyhedral region in which the current state $\tilde{x}_0$ is found. Once an index $i$ of the corresponding region is identified, then the optimal control action can be calculated using~\eqref{eq:pwa}. There exists numerous point location algorithms in literature and implemented in MPT~\cite{herceg2013mpt,kvasnica2015design}. The simplest algorithm is sequential search, which checks constraint satisfaction $Z_i \tilde{x}_0 \le z_i$ for all regions $i = 1,\dots, M$ one after another until the region containing current state $\tilde{x}_0$ is found ~\cite{ingole2015fpga}. 
\section{Embedded Realization of EMPC}
\label{sec:emb:empc}
\begin{figure} [htbp]
		\centering 
		\includegraphics[width=0.7\linewidth, height=0.6\linewidth]{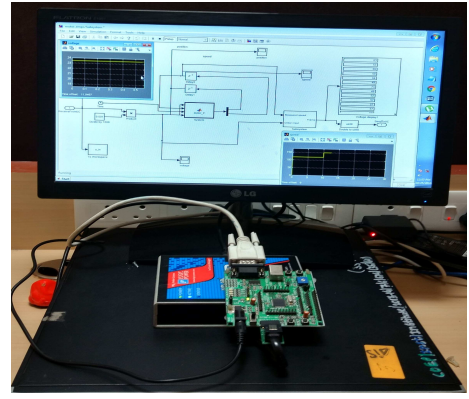}
		\caption{\pic~microcontroller SoC HIL co-simulation lab set-up used in the speed control of DC motor.} 
		\label{fig:HILlabsetup}
	\end{figure}

The robust and scalable customized data packet is developed for USART to establish communication between microcontroller and~\matlab~for HIL co-simulation. Fig.~\ref{fig:HILlabsetup} shows, the overview of HIL set-up with microcontroller and~\SIM~for the DC motor speed control using explicit MPC.
\subsection{Construction of Explicit MPC}
\label{subsec:ConstrEMPC}
A combined (motor and disturbance model) discrete time LTI model~\eqref{eq:alti} is considered to design an explicit MPC with 4 states ($x_e$), 1 input ($u$) and 1 output ($y_e$). The cost function in~\eqref{eq:cftoc}  formulated with prediction horizon of, $N = 2$, delta input penalty, $R = 1$, output penalty, $Q = 1$ and the constraints on input as $0\leq u \leq 24$. The MPC problem is constructed in open source freely available~\matlab~based Multi-Parametric Toolbox (MPT) and an explicit PWA control law~\eqref{eq:pwa} was obtained with 13 regions. Further, the sequential search algorithm for the controller is exported in the form of a library-free low-level~\cp~code. The generated~\cp~code is complies with the American National Standards Institute (ANSI) standard and can be used directly in the~\mplab~X IDE. The matrices generated (LUT) using the tool are of data type double and in XC8, the size of double has been configured as~\SI{32}{bit} in XC8 linker's memory model option.
	\begin{figure*}[htbp]
		\begin{center}
			\begin{bytefield}[endianness=little,bitheight=0.03\linewidth,bitwidth=0.075\linewidth]{13}
				\bitbox{1}{S}
				\bitbox{1}{$\hat{x}_e(t)(1)$}
				\bitbox{1}{I}
				\bitbox{1}{$\hat{x}_e(t)(2)$}
				\bitbox{1}{D}
				\bitbox{1}{$\hat{x}_e(t)(3)$}
				\bitbox{1}{C}
				\bitbox{1}{$\hat{x}_e(t)(4)$}
				\bitbox{1}{O}
				\bitbox{1}{$u_{(t-1)}$}
				\bitbox{1}{E}
				\bitbox{1}{$r(t)$}
				\bitbox{1}{P}\\
			\end{bytefield}
			\caption{Representation of data frame used to send data from~\SIM~to microcontroller.}			
			\label{fig:protocol:sm2}
		\end{center}
	\end{figure*}
\subsection{\matlab and Microcontroller Interface}
\label{subsec:communication}
USART interface module available in~\pic~has been used for asynchronously communicating with~\SIM. The data frame as prepared for single byte transfer comprised a start bit and stop bit while~\SI{8}{bit} of data is sandwiched between them.
The baud rate selected for data transfer is~\SI{115200}{bps} and the processor is running at~\SI{48}{MHz}. In the on-line phase, search algorithm needs current value of $\tilde{x_0}$ which is comprised of six variables i.e. position ($\hat{x}_e(t)(1)$), speed ($\hat{x}_e(t)(2)$), estimated speed ($\hat{x}_e(t)(3)$), estimated disturbance ($\hat{x}_e(t)(4)$), previous input ($u_{(t-1)}$) and reference ($r(t)$). All these values are sent to the microcontroller through customized data packet as shown in Fig.~\ref{fig:protocol:sm2}. For data security reasons, we need to provide the separators (S, I, D, C, O, E and P) for each of the data which has been done by appending characters before and after the variable. So, a frame has been designed in~\matlab~which after being sent will be decoded by the microcontroller and separated into the variables which are required for EMPC evaluation at each sample time. Packet size depends on the accuracy of data required for the application. Data being sent in this application is accurate up to second digit after decimal point. Five bytes of each of the six variables are being sent to the microcontroller. 
	
When frame will be sent, the microcontroller will start accumulating the data in the buffer. After the frame is completely received, then it will begin assigning the variables which would be further sent to EMPC routine for generating optimal input. This input would be later converted into another frame which can be decoded into~\matlab~for obtaining correct input signal. 
The data being sent to~\matlab~in this application is accurate up to third digit after decimal point. As maximum voltage input to motor is~\SI{24}{V},~\SI{5}{bytes} of input are being sent. Overall considering the header and terminator byte, and start and stop bits,~\SI{70}{bits} are being sent.	
\subsection{Realization of Explicit MPC on PIC Microcontroller}
\label{subsec:RealizeEMPC}
Considered microcontroller~\pic~does not provide a dynamic memory allocation~\cite{MICROCHIPxc8}. Hence, the code was tweaked so as to eliminate the need for dynamic memory allocation which was originally present in code generated by MPT. Following are the~\cp-functions used in the realization of explicit MPC, 
	\begin{verbatim}
	void HighPriorityInterrupt  IntISR(void).
	\end{verbatim} 
When an interrupt is generated, it would land here in this function (to address $0\times08$ (high priority interrupt vector address)) from where it would be redirected to {\tt{MyBufferRX}} function. This is done as the interrupt vector table has limited space.
	\begin{verbatim}
	void MyBufferRX(unsigned char).
	\end{verbatim}
In this function, the received frame is separated and data is converted to an integer from American Standard Code for Information Interchange (ASCII). 
This array is then sent to {\tt{ MPCExplicit()}} function
	\begin{verbatim}
	void MPCExplicit(double *).
	\end{verbatim}
In this function, the received array is sent to control law evaluation function i.e. {\tt{SequentialSerach()}}. The input, when received from this function, is multiplied by $1000$ and then sent to {\tt{Int2AsciiTX(unsigned long int u)}}. This is done so that the data, precise up to $3$ digits after decimal is obtained. 
	\begin{verbatim}
	double SeqentialSerach(double *X).
	\end{verbatim}
This function looks for suitable input value based upon the array of values sent to the function and returns input. This code is generated using MPT toolbox of~\matlab. 
	\begin{verbatim}
	void Int2AsciiTX(unsigned long int u).
	\end{verbatim}
This function converts integer value of the input to ASCII value and transmits data serially to~\matlab. Then ports initialization and interrupts enabling is done by following function 
	\begin{verbatim}
	void Initialise().
	\end{verbatim}
\section{HIL Co-Simulation Results}
\label{sec:Results}
This section describes the HIL co-simulation results of the implemented explicit MPC and PI controller on the microcontroller. Furthermore, the performance of explicit MPC is shown with the constraint handling, trajectory tracking, disturbance rejection and the complexity of control law is discussed. 
\subsection{Controller Performance}	
\label{subsec:result:per}
\begin{figure}[htbp]
		\centering 
		\setlength\figureheight{6cm}\setlength\figurewidth{0.42\textwidth}
		\input{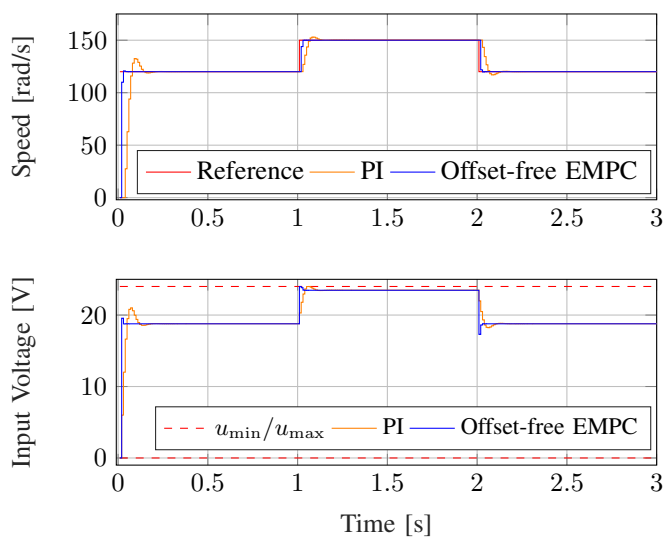} 
		\caption{Performance of PI and offset-free explicit MPC for reference tracking.} 
		\label{fig:reftrack}
	\end{figure}

Fig.~\ref{fig:reftrack} shows the output and input response, with explicit MPC and the PI controller, for the varying speed reference tracking. 
It is clearly seen that the designed MPC achieves reference speed faster compared to PI controller. Input voltage is shown in the below sub-plot, which shows that the input for both the controls are operating within the constraints.  
\begin{table*}[htbp]
	\centering
	\setlength{\tabcolsep}{3pt}
	\renewcommand*{\arraystretch}{1.3}
	\caption{Comparison of  explicit MPC control law for different prediction horizons.}
	\begin{tabular}{cccccccc}
		\toprule
		N &  $\#$ Regions &  Available ROM [\unit{kB}] &  Used ROM [\unit{kB}] & Usages [$\%$] & Available RAM [\unit{kB}]   & Used RAM [\unit{kB}]  & Usages [$\%$] \\\midrule
		2  & 13 & 32.77 & 15.94 & 48.67& 2.05& 0.33& 16\\ 
		3  & 59 & 32.77 & 30.01 & 92.00& 2.05 & 0.33& 16\\
		\bottomrule
	\end{tabular}
	\label{tab:memory}
\end{table*}

\begin{figure}[htbp]
		\centering 
		\setlength\figureheight{6cm}\setlength\figurewidth{0.42\textwidth}
		\input{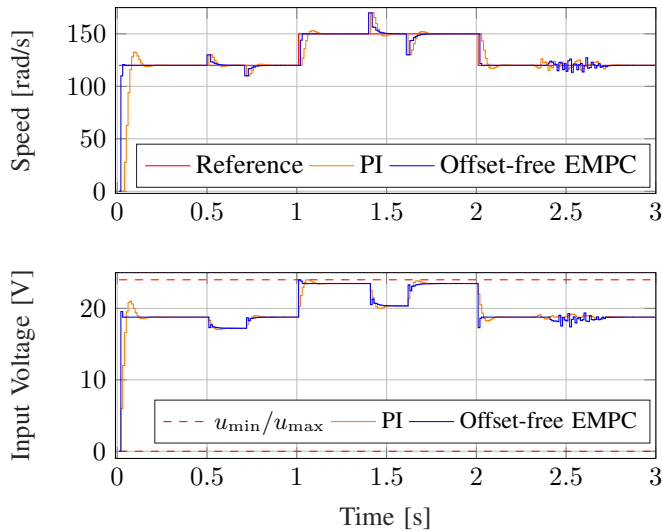} 
		\caption{Performance of PI and offset-free explicit MPC for reference tracking with disturbances.} 
		\label{fig:dist}
	\end{figure}

Further, offset-free explicit MPC is tested for the disturbance rejection. Fig.~\ref{fig:dist} shows the response of motor speed control. A disturbance of~\SI{10}{V} and~\SI{20}{V} was given at the time~\SI{0.5}{s} to~\SI{0.7}{s} and~\SI{1.4}{s} respectively. Also, to test noisy signal scenario, we added noise up to the amplitude of~\SI{8}{V} in between time~\SI{2.3}{s} to~\SI{2.7}{s}. It can be seen that EMPC performs better than PI controller and reject disturbances.  The PI controller takes more time to settle and shows oscillations which is not acceptable for the actual motor.
\subsection{Controller Complexity}	
\label{subsec:result:compl}	
The designed offset-free explicit MPC is implemented on the~\pic~microcontroller using two different prediction horizons ($N$) and subsequently, we compare controller complexity on the basis of the number of regions, memory used to store controller data, and memory used for the sequential search and communication interface code. Table~\ref{tab:memory} shows the complexity of explicit MPC controller for prediction horizon 2 and 3, for the prediction horizons larger than 3 memory exceeds the available on-chip ROM. The explicit MPC matrices ($F, g, Z$ and $z$) and the code were saved on the ROM and other data need for the implementation was saved on RAM. It can be seen from the table that $N = 2$ needs 43.3$\%$ less memory as compared to the $N = 3$ but in the form or performance quality both the controllers give almost same output, that's why explicit controller with $N = 2$ can be the appropriate choice for this application.
\section{Conclusions}
\label{sec:conclusion}
This paper focuses on the implementation of embedded MPC on a low-cost~\pic~microcontroller. The designed offset-free explicit MPC for the speed control of DC motor with disturbance modeling approach using multi-parametric toolbox which generates the low-level~\cp.It is tweaked and modified so as to run on the microcontroller platform. The results obtained indicate that the EMPC solution outperforms the conventional PI controller with respect to performance and controller efforts to track reference with reasonable utilization of program memory. The HIL co-simulation results obtained from MPT are highly satisfactory and reflects the effectiveness of proposed embedded solution for motor speed control and the potential embedded applications. 
\section*{Acknowledgment}
We gratefully acknowledge the support from R \& D center of the COEP. Deepak Ingole would like to thank for a financial contribution from the European Research Council (ERC) under the European Unions Horizon 2020 research and innovation program (grant agreement No 646592 MAGnUM project). 
\balance
\bibliographystyle{IEEEtran}
\bibliography{ifacconf} 
\end{document}